\documentclass[fleqn,10pt]{wlscirep}
\usepackage[utf8]{inputenc}
\usepackage[T1]{fontenc}
\title{Passive symmetry breaking of the space-time propagation in cavity dissipative solitons}

\author[1]{Idan Parshani}
\author[1,2]{Leon Bello}
\author[1]{Mallachi-Elia Meller}
\author[1,*]{Avi Pe'er}
\affil[1]{Department of Physics and BINA Institute of Nanotechnology, Bar-Ilan University, Ramat-Gan 52900, Israel}
\affil[2]{Department of Electrical and Computer Engineering, Princeton University, Princeton, NJ}

\affil[*]{Correspondence email address: {avi.peer@biu.ac.il}}


\begin{abstract}
Dissipative solitons are fundamental wave-pulses that preserve their form in the presence of periodic loss and gain. The canonical realization of dissipative solitons is Kerr-lens mode locking in lasers, which delicately balance nonlinear and linear propagation in both time and space to generate ultrashort optical pulses. This linear-nonlinear balance dictates a unique pulse energy, which cannot be increased (say by elevated pumping), indicating that excess energy is expected to be radiated in the form of dispersive or diffractive waves. Here we show that Kerr-lens mode-locked lasers can overcome this expectation. Specifically, by breaking the spatial symmetry between the forward and backward halves of the round-trip in a linear cavity, the laser can modify the soliton in space to incorporate the excess energy. Increasing the pump power leads therefore to a different soliton solution, rather than to dispersive / diffractive loss. We predict this symmetry breaking by a complete numerical simulation of the spatio-temporal dynamics in the cavity, and confirm it experimentally in a Kerr-lens mode-locked Ti:Sapphire laser with quantitative agreement to the simulation. The simulation opens a window to directly observe the nonlinear space-time dynamics that molds the soliton pulse, and possibly to optimize it. 
\end{abstract}
\begin{document}

\flushbottom
\maketitle
%
%
\thispagestyle{empty}

\section*{Introduction}

Solitons are waves with exceptional stability properties, which appear in many areas of physics. They are a fascinating non-linear phenomenon where wave propagation becomes dispersion-less and diffraction-less. We focus on \emph{dissipative} solitons, which develop in cavities with both gain and loss, that preserve their spatio-temporal shape from one round trip to the next, but not within the round trip.

Solitons are ubiquitous in all areas of physics, ranging from hydrodynamics through condensed matter to optics \cite{Soliton_Wave}. Specifically, optical solitons have important implications to fiber optics and communications \cite{haus_solitons_in_comm_1996, mollenauer_solitons_in_fibers_2006}, ultra-fast optics and mode-locked lasers \cite{ryczkowski_real-time_2018, haus_mode-locking_2000, t_brabec_kerr_1992, wright_mechanisms_2020}. In the field of ultrafast optics, solitons have been harnessed for the generation of ultrashort pulses and frequency combs in mode-locked lasers \cite{cundiff-frequency_comb_colloquium_2003, f_x_kartner_femtosecond_2005}, passive fiber resonators \cite{peng_real-time_2018,Leo_2010,Oktem_2010,Wu:21}, and micro-resonators \cite{zhang_optomechanical_2021,Herr_2013,Yi_2015,Brasch_2016,Obrzud_2017}.

\begin{figure}[h]
    \centering
    \includegraphics[width = 0.75\textwidth]{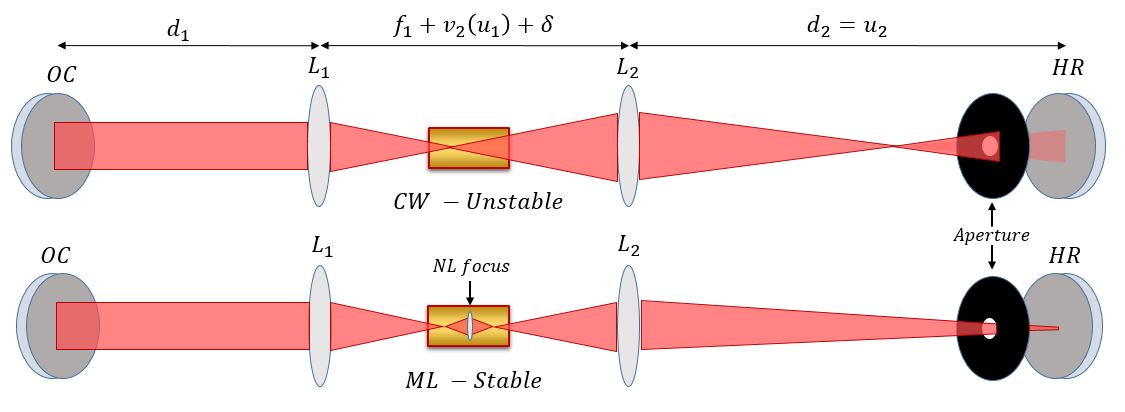}
    \caption{The soliton model of KLM: The Kerr effect in the gain medium acts to focus the beam (red) inside the cavity, mitigating diffraction losses for pulses with high peak power. \textbf{(top)} Linear beam propagation inside the cavity \textbf{(bottom)} Beam propagation with the Kerr lens for pulsed operation. \textbf{$d_{12}$} are the distances in the free propagation arms of the cavity, between the end mirrors (denoted by \textbf{OC} and \textbf{HR}) and the focusing elements (denoted \textbf{$L_{1,2}$}) with focal lengths $f_{1,2}$, respectively. The total distance between the focusing elements deviates from  the imaging condition by $\delta$, which quantifies the cavity stability (or deviation from it). The aperture, which illustrates the role of the diffraction losses in the cavity, can be an actual element or an effective aperture induced by the overlap between the cavity mode and the pump beam in the gain medium.} 
    \label{fig:ml_and_cw}
\end{figure}

In optics, Kerr-lens mode-locking (KLM) is considered the canonical example of soliton generation, which is widely explored \cite{ryczkowski_real-time_2018}. Specifically, KLM is a passive mode-locking technique for producing ultra-short pulses \cite{herrmann_theory_1994, t_brabec_kerr_1992, t_brabec_hard-aperture_1993}, where a third-order (Kerr) non-linear lensing effect, where the gradient of the spatial profile, together with the intensity-dependent self-phase modulation effect, create an effective non-linear lens. Combined with an effective hard-aperture in the cavity, this acts to overcome the diffraction losses from the hard-aperture that low-power CW experiences. This nonlinear lens stabilizes pulses of high peak power by focusing them through the aperture, thereby reducing the loss for pulses \cite{kurtner_mode-locking_1998, matsko_mode-locked_2011, haus_mode-locking_2000}. These stable Kerr pulses are solitons, and are only stable for the specific intra-cavity pulse power that generates the required focal length for the Kerr-lens to precisely counteract the diffraction loss of the cavity \cite{herrmann_theory_1994, haus_theory_1975, Liu2019, Jung1995, ippen_principles_1994, Kartner1996, Haus1996}, as illustrated in figure \ref{fig:ml_and_cw}. In this work, we show that this strict power condition can be alleviated by taking advantage of a new effect in lasers with multiple Kerr-lens interactions.

The spatial profile of a laser beam across the cavity is normally dictated by cavity-stability analysis, by requiring the cavity mode to reproduce its shape after every round trip. 
In a linear cavity, the propagation in the forward and backward directions through the cavity is inherently symmetric, due to the inversion symmetry of all optical elements and the exact inversion of the phase-front on the end mirrors \cite{siegman_lasers_1986}. 
The standard model of KLM in a linear cavity assumes this symmetry also for the nonlinear Kerr-lens that stabilizes the cavity mode for pulsed operation (see figure \ref{fig:ml_and_cw}), which requires a specific peak-power and pulse shape to form the required focal length of the Kerr lens, and hence - a soliton. 
Once formed, the pulse energy is fixed, and any additional power in the cavity (due to elevated pumping, for example) cannot increase the circulating energy, but rather be lost to amplified-spontaneous emission (ASE), or lead to additional continuous-wave (CW) lasing, or additional pulses in the oscillation \cite{coen_modeling_2012, meller_mode-locking_2017}.

\begin{figure}
    \centering
    \includegraphics[width=0.75\textwidth]{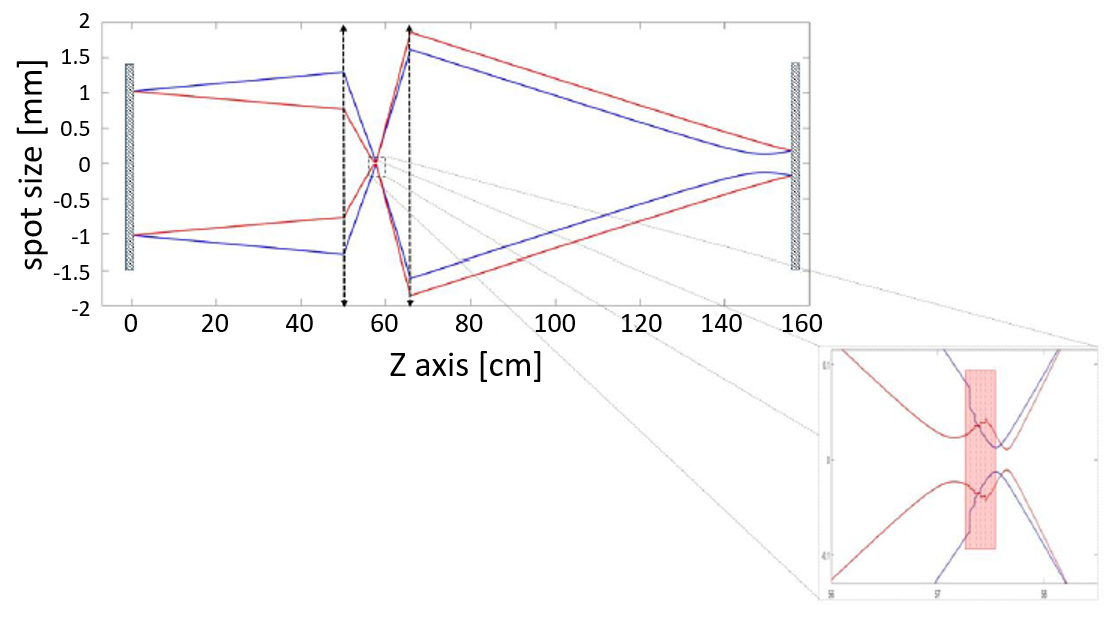}
    \caption{Symmetry-breaking between the forward and backwards halves of the cavity round trip (simulation): The oscillator breaks the propagation symmetry in order to employ higher pulse power and energy. The figure shows the numerically simulated beam width during the entire round-trip in the cavity, where the forward half (blue) is different from the backwards half (red), exhibiting different focusing power through each. \textbf{Inset:} zoom-in on the beam profile within the nonlinear Kerr medium}
    \label{fig:asym_prop}
\end{figure}

In a linear cavity the beam interacts with the Kerr medium twice per round-trip, once in the forward direction and once backwards (as opposed to a single interaction in a ring cavity). This dual nonlinearity opens an additional freedom to optimize the soliton solution. We show quite generally, that the pulses in a linear cavity exploit this freedom to mitigate the diffraction losses by breaking the cavity symmetry between forward and backwards propagation, which is a major difference from KLM in ring cavities \cite{dunlop_master_1997, c_j_chen_self-starting_1995}. As outlined in figure \ref{fig:asym_prop}, the pulse generates a stronger non-linear lens in the forward half of the round trip, which leads to over-focusing in the forward direction at the expense of a weaker focus in the backwards half (due to a larger beam area and lower intensity). Consequently, only after a complete round-trip will the pulse repeat its shape and spatial profile, but not between the halves of the round-trip. Note that this is purely a spatial effect and not temporal (as one may assume). Specifically, the symmetry breaking is not due to the different chromatic dispersion elements between the two interactions, but rather a power-dependent variation of the spatial beam that allows the soliton solution to accommodate more energy.  

We identified this effect theoretically and then confirmed it experimentally with a home-brewed KLM Ti:Sapphire cavity. The theoretical prediction originated from a numerical simulation tool that we developed to observe the complete spatio-temporal dynamics and evolution of a pulse in the cavity, under rather general assumptions. Previously, this tool was used for demonstrating the dynamical loss mechanism in KLM lasers \cite{parshani_diffractive_2021}.

It is known that Kerr nonlinearity can break the symmetry between different oscillation modes that are originally symmetric. For example, the symmetry between the two polarizations in micro-ring cavities is broken by Kerr \cite{xu_spontaneous_2021}, and the clockwise and counter-clockwise oscillations modes cannot co-exist in ring fibers with Kerr nonlinearity \cite{del_bino_symmetry_2017, xu_spontaneous_2021, hendry_spontaneous_2018}. Here we show that the symmetry can be broken even within a single mode of oscillation, when the Kerr-lens is localized in more than a single point. We emphasize that the symmetry breaking here is different from the spontaneous symmetry breaking of a Hopf bifrucation \cite{SIGLER2005305, akhmediev1993novel, chu1993soliton, nguyen2020reversible}. We are not concerned with the appearance of two different modes of operation that break the system symmetries, or a change in the system’s stability like a Hopf bifurcation, but rather the evolution of a single mode, whose symmetry evolves as we change the excitation parameters. Specifically, the two propagation directions in the cavity, which are symmetric at low pump power become asymmetric as the pump power is increased. These two directions are not modes of the laser, but rather two halves of one round-trip mode. Consequently, this symmetry breaking is not "spontaneous", but rather a continuous deterministic modification of the oscillator mode (which is not bi-stable).

Our simulation offers a powerful utility for nonlinear optical dynamics with KLM, since it exposes the complete spatio-temporal propagation of the pulse \emph{within the nonlinear medium}, where experimental probes do not exist. It simulates the evolution of the ultrafast laser oscillation on both fast and slow time-scales. The fast time-scale captures the dynamics within a single round-trip, inflicted by the Kerr lens on the temporal envelope of the pulse and the temporal beam profile (the time dependent waist and phase front). The slow time-scale represents the evolution from one round-trip to the next, simulating the convergence towards steady state due to the nonlinear gain dynamics in the laser. This provides the critical capability to observe the nonlinear dynamical evolution on all relevant time-scales, allowing us to \textit{quantitatively} explore novel concepts of KLM prior to experimental implementation. 

Since Kerr-lens mode-locked lasers operate in a single spatial mode that is nearly Gaussian, and since the Kerr-lens mechanism strongly drives the oscillation towards single-mode operation, our simulation assumes a Gaussian beam profile. Specifically, for the KLM rto generate an effective nonlinear lens, the gradient of the spatial power has to mimic a lens, i.e. a single-mode, which is strongly peaked around the center, well approximated by a Gaussian profile. In addition, the Gaussian approximation has the added benefit of simplifying the calculations greatly, since it enables to employ ABCD propagation, which is a standard procedure in cavity analysis. ABCD propagation of Gaussian beams, which is derived directly from the paraxial Maxwell-Bloch equations, is well-established and widely used. It allows us to propagate the complex beam parameter through the cavity using only the ABCD matrix of the cavit.

Since our system is dynamic and nonlinear, our non-linear optical elements are changing in time together with the beam profile and beam power. We employ a time-dependent ABCD matrix $M_n(t)$ to represent the cavity propagation of round-trip $n$ for every time-bin $t$ of the pulse. Repeating the propagation from one round-trip to the next, while updating the ABCD matrix according to the laser evolution allows to observe the complete cavity dynamics from initiation to steady state oscillation. Although other numerical analyses of KLM have used the temporal ABCD formalism before \cite{MAGNI1993348,coen_modeling_2012, salin_mode_1991,yoo_byung_duk_numerical_2005,Juang1997,t_brabec_kerr_1992,Cerullo1994,Henrich1997}, they all assumed a Kerr interaction localized at a single point and were all aimed directly at finding the steady state oscillation, not at the dynamical cavity evolution.

\begin{figure}
    \centering
    \includegraphics[width=0.75\textwidth]{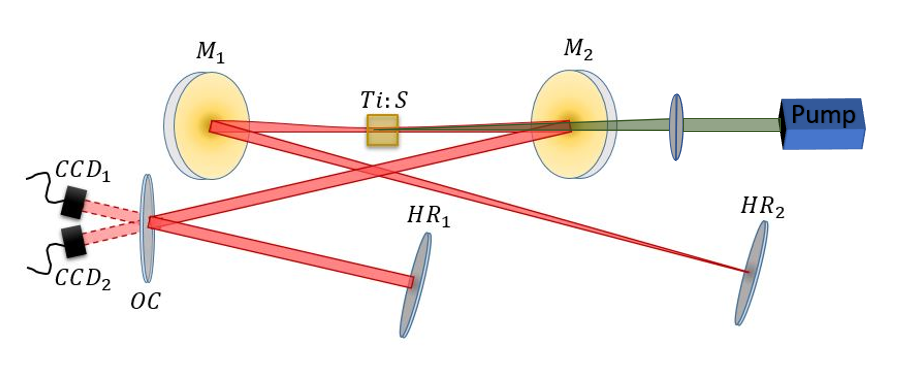}
    \caption{Experimental configuration: We employ a standard X-folded four-mirror cavity. \textbf{$M_{1,2}$} are spherical mirrors with focus $f\! =\! 75 \rm mm$. \textbf{$HR_{1,2}$} are the highly reflecting end mirrors (flat). \textbf{OC} is the output coupler used to probe the beam twice in each round-trip. A pair of BK7 prisms is deployed to compensate for intra-cavity dispersion (not shown). The gain medium (and Kerr medium) is a Brewster-cut 3mm long Ti:Sapphire crystal (0.25\% doped), pumped by a 4W-6W laser at $532nm$ (Verdi V18 by Coherent). The generated pulses have a repetition rate of $67\rm MHz$ and a duration of roughly $50 \rm fs$.}
    \label{fig:experimental_scheme}
\end{figure}

In section \ref{sec:results} below details the experimental observations of symmetry breaking within the cavity for output power, beam profile and laser-threshold, and highlights the quantitative agreement of the experimental results with the numerical predictions. A review of the numerical simulation is provided in section \ref{sec:methods}. 

\begin{figure}
    \centering
    \includegraphics[width=0.5\textwidth]{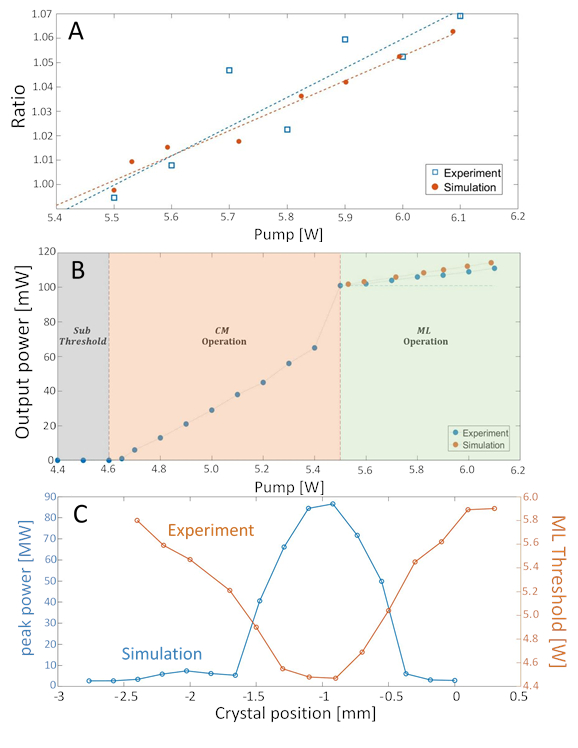}
    \caption{\textbf{Results: A. Ratio of intra-cavity beam width between the forward and backwards directions.} Blue rectangles - measured, as captured on the two cameras (CCD1 and CCD2). Orange circles - numerical simulation. the asymmetry ratio increases with the pump power in accordance with the increase of the total output power and with the deviation from the soliton assumption. \textbf{B. Output power as function of input pump power}, measured experimentally (blue). We observe three regions. \textbf{(1)} Below the lasing threshold, where the output power is low. \textbf{(2)} Above the CW lasing threshold, where the output power varies linearly with the pump. \textbf{(3)} Above the ML lasing threshold, where a linear increase is still observed, but with a lower slope, deviating from the soliton assumption. For comparison, we also show the simulation predicted output power (orange, fitted only for the ML threshold), which predicts nearly the same power slope as the measurement.\textbf{ C. Optimal crystal position for ML efficiency} - experiment and numerical simulation. \textbf{Blue} - the numerically calculated ML peak-power, and orange - the experimental ML threshold, as a function of the crystal offset from the focal point of the short arm's lens towards the long arm of the cavity (see figure \ref{fig:ml_and_cw} for definition). Interestingly, the optimal crystal position is offset \textit{in the opposite direction} from the conventional expectation assuming a soliton mode.}
    \label{fig:results}
\end{figure}

\section*{Results}

\label{sec:results}
In our experiment, we employ a standard Ti:Sapphire oscillator in an X-folded cavity \cite{yefet_mode_2013, yefet_review_2013}, as shown in figure \ref{fig:experimental_scheme}. In order to observe the intra-cavity beam in both forward and backwards directions, a planar output coupler (reflectivity $R = 0.98$) folds the cavity configuration near one of the focusing mirrors and couples out a fraction of the intra-cavity beam in both directions to be imaged on a CCD camera. The results show symmetry breaking with quantitative agreement to the numerical simulation. As illustrated in figures \ref{fig:results}A,B the symmetry breaking allows the pulse power to increase with the pump power beyond the soliton level. 

The distance between the mirrors is offset by $\delta$ from the geometrical imaging condition, corresponding to different stable or unstable ray configurations. Our cavity operates slightly outside the stability zone, which induces diffraction losses for CW operation, while for high peak-intensity pulses, the Kerr-lens counteracts the diffraction losses and pushes the cavity back into spatial stability. The strength of the effective fast saturable absorber that is induced by the Kerr-lens 
can be controlled by adjusting $\delta$, i.e. moving further away from the stability zone.

Our analysis (figure \ref{fig:asym_prop}) shows a broad regime outside the stability zone and above the oscillation threshold, where the laser changes its temporal profile such that it induces different optical powers in the Kerr medium through the two otherwise identical halves of the round-trip. This asymmetry shows a quantifiable signature - the beam waist at the same location changes considerably between the forward and backwards direction. All of these findings were first predicted by the numerical simulation and only later verified experimentally.

Figure \ref{fig:results}\textbf{A} shows the ratio between the measured widths (FWHM) of the forward and backward beams on the CCD, while scanning the pump power. Clearly the beam size asymmetry increases linearly with the pump, in good agreement with the numerical predictions. In accordance, the output power of the laser (figure \ref{fig:results}\textbf{B}) also shows a linear increase with pump power, demonstrating that the laser find an efficient soliton solution as the pump power is increased and the diffractive/dispersive losses are mitigated.
The measured disagreement between our (rather simple) model and the full experiment in \ref{fig:results}\textbf{A+B} is well within the experimental and numerical error-range, and amounts to $12\%$ at most. We emphasize that this rather quantitative agreement relies on  a \emph{single} fit parameter for all results (the ML threshold). The small discrepancies between the experiment and the model can be accounted for by additional parameters, such as the total cavity GVD, intrinsic losses, etc, but  these parameters are quite hard to measure accurately.

It is interesting to note that the simulation directly contradict a symmetric cavity assumption. Specifically, the simulation predicts that the maximum output power (at a fixed pump) is achieved when the crystal is offset from the focal point towards the short arm by $\sim1\rm mm$ \cite{yoo_byung_duk_numerical_2005}. This optimal offset cannot be reconciled with the assumption of symmetric propagation in both directions in the cavity, which would infer that the lens position can only be offset \textit{in the opposite direction} towards the long arm. The experimental measurement of the pump threshold for ML at different crystal positions shows a clear optimum with a width that agrees well with the simulation (see figure \ref{fig:results}C). Although the experimental uncertainty in the exact offset position of the crystal ($\sim\!1$mm) prevents unequivocal verification of the optimum location, this agreement is an additional support for the validity of the simulation.

\section*{Conclusions}
In summary, we uncovered through numerical analysis, and verified experimentally, a passive symmetry breaking effect in KLM lasers that mitigates the diffractive losses when multiple Kerr-lens interaction exist in the cavity. Namely, the spatial symmetry between identical parts of the cavity can be broken when the Kerr lens interaction occurs more than once during the round-trip, as is the case in a linear cavity for the forward and backwards propagation. This symmetry breaking is a fundamental freedom for dissipative solitons in space that was not explored before. It allows the laser to optimize the soliton solution in space, and enhance its power efficiency beyond a single soliton solution \cite{herrmann_theory_1994}. It is attractive to explore cavities with even more Kerr-lens interactions \cite{meller_mode-locking_2017}, where this freedom may modify the soliton solution much further.

In addition to understanding the internal dynamics of KLM lasers, which may be surprising and non-intuitive, our work has important implications to the design of KLM oscillators. Specifically, the inclusion of several non-linear lensing locations in the cavity provides the soliton with an additional degree of freedom that can lead to efficient utilization of the pump-power, while preserving its soliton nature. In standard soliton theory, the soliton solution exists only for a limited range of pulse powers, and driving more power into the system (for example, by increasing the pump-power), would normally not increase the soliton power but would rather excite other, often unwanted, modes - such as CW, or additional parasitic pulses. The additional non-linear lens allows the soliton solution to exist in a much wider range of pulse powers.

\section*{Methods}
\label{sec:methods}
The concept and operation procedure of the numerical simulation is reviewed hereon. A thorough detail of the simulation, along with additional verification tests will be provided in a future publication.

Our simulation calculates the dynamical evolution of the pulse in the cavity in both space and time, and shows its dynamical properties as well as its final steady-state. We focus on the hard-aperture KLM regime, where diffraction losses are the dominant effect.
We approximate the spatial mode of the laser to be a single transverse Gaussian $TEM_{00}$ mode, whose waist is intensity-dependent and varies in time due to the Kerr-lens (which is time-dependent).

\begin{subequations}
    \begin{align}
    E(x, z) &= \exp{\left ( -i \frac{\pi x^2}{q(z, t) \lambda} \right )} \\
    \frac{1}{q(z, t)} &= \frac{1}{R(z, t)} - i \frac{\lambda}{\pi w(z, t)^2}
    \end{align}
\end{subequations}

where $\lambda$ is the wavelength inside the element, $z$ is the propagation distance, $w(z)$ is the beam width at point $z$ and $R$ is the beam's radius of curvature.
Under this Gaussian approximation, the spatial amplitude profile of the beam during the $n$-th round trip is fully represented by a time-dependent complex beam parameter $q_{n}(t)$. 

Following the derivation in Siegman \cite{siegman_lasers_1986}, we solve the Huygens' integral for a Gaussian mode impinging on an element whose paraxial evolution is described by an ABCD matrix. Solving the resulting Gaussian integral, gives yet another Gaussian mode. This directly relates the input and output Gaussian beams according to,

\begin{equation}
    E'(x, z) = \frac{1}{\sqrt{A + n_1 B/q(z, t)}} \exp{\left ( -i \frac{\pi x^2}{q'(z, t) \lambda} \right )}
\end{equation}

Following that discussion, we employ a lumped-element, discrete-time approach, which lends itself naturally to free-space laser analysis, like the system we study in this work. In that approach, the beam in the output plane of each element is related to the beam in the input plane by an ABCD matrix.

\begin{equation}
    q'(z, t) = \frac{A q(z, t) + B}{C q(z, t) + D}
\end{equation}

which follows the standard cavity propagation using ABCD analysis \cite{siegman_lasers_1986}. The total scattering matrix for the cavity is obtained by multiplying the matrices of all constituting elements in the cavity, relating the complex beam-parameter between one roundtrip and the next one.

This assumption of a single spatial-mode is legitimate for KLM oscillators in free-space, which are strongly driven by the hard aperture towards a single spatial mode \cite{yefet_review_2013}. Specifically, higher transverse modes are not able to properly induce the Kerr lens and are strongly suppressed by the diffraction losses.

\begin{figure}
    \centering
    \includegraphics[width=0.6\textwidth]{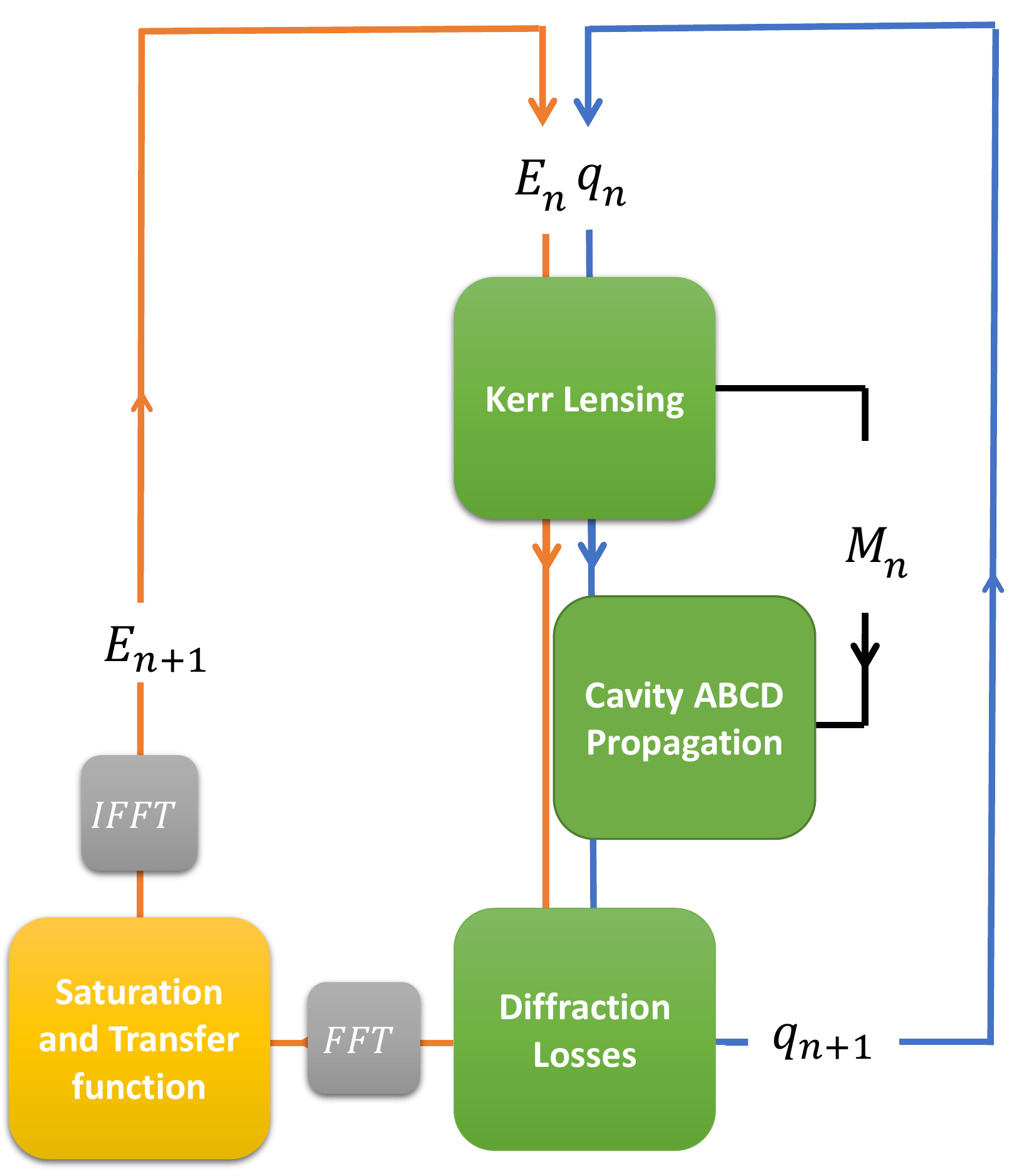}
    \caption{Flow diagram of the simulation. In each round-trip $n$, the simulation maintains two vectors - $E_n(t)$, the instantaneous field envelope of the pulse and $q_n(t)$, the instantaneous complex beam parameter.
    Using those two vectors, we can calculate the instantaneous non-linear lens based on the instantaneous intensity distribution, which can then be combined into the complete ABCD matrix $M_n(t)$ of the cavity round-trip, which allows to propagate the beam and calculate the new complex beam parameter $q_{n+1}(t)$. The instantaneous diffraction losses during the round-trip can then be calculated based on the new complex beam parameter and the assumed aperture function. Finally, gain saturation, finite gain bandwidth and cavity dispersion are applied in frequency domain to the field, providing the field envelope $E_{n+1}(t)$ for the next round-trip.}
    \label{fig:simulation_flow}
\end{figure}

Note that the ABCD matrix $M_n(t)$ of the $n$-th round-trip is time dependent as well, since it includes the intensity-dependent Kerr-lens, which varies according to the temporal intensity of the intra-cavity pulse, as well as from one round-trip to the next (due to the gain evolution). 
We calculate therefore the complex field envelope on two time-scales, a ``slow'' time-scale, which accounts for the variations from one round-trip to the next, and a ``fast'' time-scale, which measures the intra-cavity evolution within the round-trip due to the Kerr-lens effect. Our simulation makes no explicit definition of the carrier frequency, which is completely arbitrary, and simulates only the complex temporal envelope of the pulse. The simulation flow is outlined in figure \ref{fig:simulation_flow}, applying dispersion and linear gain/loss in frequency-domain, while saturation and the Kerr-effect are applied in the time-domain, where they are more naturally described.

The simulation accepts several parameters that reflect the known (or measured) properties of the laser in question - the nonlinear Kerr coefficient, the net dispersion of the cavity, the gain bandwidth (assuming for now a Gaussian gain spectrum), the small-signal gain and loss, gain saturation parameter, gain spatial profile (pump-mode width at the gain medium) and the loss function. The aperture can be located anywhere in the cavity. For example, placing the aperture near one of the end mirrors will simulate a hard aperture, whereas an aperture near the gain crystal simulates a soft aperture, where the diffraction losses are due to the spatial mismatch between the cavity mode and the gain-profile (the pump spot size in the crystal). 

In each step, we update two major vectors: the pulse field envelope $E_{n}(t)$ and the complex beam parameter $q_{n}(t)$ in each round trip $n$. Initially, the field envelope and complex beam parameter are taken as low-intensity random noise $(E_0(t), q_0(t))$, that represent the spontaneous emission seed. They are then propagated repeatedly through four modules (see figure \ref{fig:simulation_flow}), simulating the important steps in the pulse evolution: Kerr lensing and self-phase modulation, cavity propagation, diffraction losses (hard aperture), gain and dispersion. This process is repeated until the intra-cavity pulse profile stabilizes. The simulation parameters were matched to our experimental apparatus like the distance between the mirrors and the lenses and the thickness of the gain medium, but clearly the parameters can be matched to other systems as well. 

The cavity propagation employs a time-dependent ABCD matrix $M_n(t)$ that incorporates the non-linear lens along with the other linear optical elements in the cavity. The focal power of the non-linear Kerr-lens is calculated according to
\begin{equation}
    \frac{1}{f_{n}(t)}= 8n_{2}\frac{\rm{d}{P}_{n}(t)^{2}}{\pi{w_{n}}(t)^{4}}
\end{equation}
based on the intra-cavity power $P_n(t)$, the beam width $w_n(t)$ (from the complex beam parameter $q_n(t)$), the non-linear refractive index $n_{2}$ and  and the thickness of the Kerr medium $d$.  

Finally, we calculate the the field envelope for the next round trip $E_{n+1}(t)$ by calculating the time dependent gain, loss and group-velocity dispersion. The time-dependent diffraction losses due to the aperture are calculated in time domain, whereas the gain bandwidth, dispersion and linear losses are computed in frequency domain, where they are efficiently represented with a transfer function of the spectral gain and dispersion profile. To account for gain saturation, we calculate the mean power over the roundtrip $\bar{P}_{n}$ and compute the gain saturation $g_{n}=1/(1+\bar{P_{n}}/P_{ss})$, where $P_{ss}=2.6\rm W$ is the saturation power, calculated from the transition cross section and the upper-state lifetime of Ti:Sapphire \cite{rp_photonics_sat_power}. The gain dynamics is assumed to be slow compared to the pulse time scale, which is well validated for femtosecond pulses in a CW-pumped Ti:Sapphire oscillator. Specifically, our simulation incorporates gain depletion in each round-trip that is proportional to the total pulse energy of that round-trip. We assume that gain replenish occurs slowly between cavity round-trips, but neglect the dynamics within the ultrashort pulse itself, which is well-justified on the femtosecond time-scale of the pulse.

\section*{Data availability}
All data generated or analysed during this study are included in this published article 

\noindent

\bibliography{main.bib}

\begin{thebibliography}{10}
\urlstyle{rm}
\expandafter\ifx\csname url\endcsname\relax
  \def\url#1{\texttt{#1}}\fi
\expandafter\ifx\csname urlprefix\endcsname\relax\def\urlprefix{URL }\fi
\expandafter\ifx\csname doiprefix\endcsname\relax\def\doiprefix{DOI: }\fi
\providecommand{\bibinfo}[2]{#2}
\providecommand{\eprint}[2][]{\url{#2}}

\bibitem{Soliton_Wave}
\bibinfo{author}{Remoissenet, M.}
\newblock \emph{\bibinfo{title}{Waves called solitons : concepts and
  experiments}} (\bibinfo{publisher}{Springer}, \bibinfo{address}{Berlin},
  \bibinfo{year}{1994}).

\bibitem{haus_solitons_in_comm_1996}
\bibinfo{author}{Haus, H.~A.} \& \bibinfo{author}{Wong, W.~S.}
\newblock \bibinfo{journal}{\bibinfo{title}{Solitons in optical
  communications}}.
\newblock {\emph{\JournalTitle{Reviews of modern physics}}}
  \textbf{\bibinfo{volume}{68}}, \bibinfo{pages}{423} (\bibinfo{year}{1996}).

\bibitem{mollenauer_solitons_in_fibers_2006}
\bibinfo{author}{Mollenauer, L.~F.} \& \bibinfo{author}{Gordon, J.~P.}
\newblock \emph{\bibinfo{title}{Solitons in optical fibers: fundamentals and
  applications}} (\bibinfo{publisher}{Elsevier}, \bibinfo{year}{2006}).

\bibitem{ryczkowski_real-time_2018}
\bibinfo{author}{Ryczkowski, P.} \emph{et~al.}
\newblock \bibinfo{journal}{\bibinfo{title}{Real-time full-field
  characterization of transient dissipative soliton dynamics in a mode-locked
  laser}}.
\newblock {\emph{\JournalTitle{Nature Photonics}}}
  \textbf{\bibinfo{volume}{12}}, \bibinfo{pages}{221--227},
  \doiprefix\url{10.1038/s41566-018-0106-7} (\bibinfo{year}{2018}).

\bibitem{haus_mode-locking_2000}
\bibinfo{author}{Haus, H.~A.}
\newblock \bibinfo{journal}{\bibinfo{title}{Mode-{Locking} of {Lasers}}}.
\newblock {\emph{\JournalTitle{IEEE Journal on Selected Topics in Quantum
  Electronics}}} \textbf{\bibinfo{volume}{6}} (\bibinfo{year}{2000}).

\bibitem{t_brabec_kerr_1992}
\bibinfo{author}{T.~Brabec, P. F.~C., Ch.~Spielmann} \&
  \bibinfo{author}{Krausz, F.}
\newblock \bibinfo{journal}{\bibinfo{title}{Kerr lens mode locking}}.
\newblock {\emph{\JournalTitle{Optics Letters}}} \textbf{\bibinfo{volume}{17}}
  (\bibinfo{year}{1992}).

\bibitem{wright_mechanisms_2020}
\bibinfo{author}{Wright, L.~G.} \emph{et~al.}
\newblock \bibinfo{journal}{\bibinfo{title}{Mechanisms of spatiotemporal
  mode-locking}}.
\newblock {\emph{\JournalTitle{Nature Physics}}}
  \doiprefix\url{10.1038/s41567-020-0784-1} (\bibinfo{year}{2020}).
\newblock \bibinfo{note}{Publisher: Springer Science and Business Media LLC}.

\bibitem{cundiff-frequency_comb_colloquium_2003}
\bibinfo{author}{Cundiff, S.~T.} \& \bibinfo{author}{Ye, J.}
\newblock \bibinfo{journal}{\bibinfo{title}{Colloquium: Femtosecond optical
  frequency combs}}.
\newblock {\emph{\JournalTitle{Reviews of Modern Physics}}}
  \textbf{\bibinfo{volume}{75}}, \bibinfo{pages}{325} (\bibinfo{year}{2003}).

\bibitem{f_x_kartner_femtosecond_2005}
\bibinfo{author}{F.~X.~Kärtner, S. T.~C., E. P.~Ippen}.
\newblock \emph{\bibinfo{title}{Femtosecond {Optical} {Frequency} {Comb}:
  {Principle} {Operation} and {Applications}}}
  (\bibinfo{publisher}{USA:Springer}, \bibinfo{year}{2005}).

\bibitem{peng_real-time_2018}
\bibinfo{author}{Peng, J.} \emph{et~al.}
\newblock \bibinfo{journal}{\bibinfo{title}{Real-time observation of
  dissipative soliton formation in nonlinear polarization rotation mode-locked
  fibre lasers}}.
\newblock {\emph{\JournalTitle{Communications Physics}}}
  \textbf{\bibinfo{volume}{1}}, \bibinfo{pages}{20},
  \doiprefix\url{10.1038/s42005-018-0022-7} (\bibinfo{year}{2018}).

\bibitem{Leo_2010}
\bibinfo{author}{Leo, F.} \emph{et~al.}
\newblock \bibinfo{journal}{\bibinfo{title}{Temporal cavity solitons in
  one-dimensional kerr media as bits in an all-optical buffer}}.
\newblock {\emph{\JournalTitle{Nature Photonics}}}
  \textbf{\bibinfo{volume}{4}}, \bibinfo{pages}{471--476},
  \doiprefix\url{https://doi.org/10.1038/nphoton.2010.120}
  (\bibinfo{year}{2010}).

\bibitem{Oktem_2010}
\bibinfo{author}{Oktem, B.}, \bibinfo{author}{Ülgüdür, C.} \&
  \bibinfo{author}{Ömer Ilday, F.}
\newblock \bibinfo{journal}{\bibinfo{title}{Soliton{\textendash}similariton
  fibre laser}}.
\newblock {\emph{\JournalTitle{Nature Photonics}}}
  \textbf{\bibinfo{volume}{4}}, \bibinfo{pages}{307--311},
  \doiprefix\url{https://doi.org/10.1038/nphoton.2010.33}
  (\bibinfo{year}{2010}).

\bibitem{Wu:21}
\bibinfo{author}{Wu, Y.}, \bibinfo{author}{Pourbeyram, H.},
  \bibinfo{author}{Christodoulides, D.~N.} \& \bibinfo{author}{Wise, F.~W.}
\newblock \bibinfo{journal}{\bibinfo{title}{Weak beam self-cleaning of
  femtosecond pulses in the anomalous dispersion regime}}.
\newblock {\emph{\JournalTitle{Opt. Lett.}}} \textbf{\bibinfo{volume}{46}},
  \bibinfo{pages}{3312--3315}, \doiprefix\url{10.1364/OL.430926}
  (\bibinfo{year}{2021}).

\bibitem{zhang_optomechanical_2021}
\bibinfo{author}{Zhang, J.} \emph{et~al.}
\newblock \bibinfo{journal}{\bibinfo{title}{Optomechanical dissipative
  solitons}}.
\newblock {\emph{\JournalTitle{Nature}}} \textbf{\bibinfo{volume}{600}},
  \bibinfo{pages}{75--80}, \doiprefix\url{10.1038/s41586-021-04012-1}
  (\bibinfo{year}{2021}).

\bibitem{Herr_2013}
\bibinfo{author}{Herr, T.} \emph{et~al.}
\newblock \bibinfo{journal}{\bibinfo{title}{Temporal solitons in optical
  microresonators}}.
\newblock {\emph{\JournalTitle{Nature Photonics}}}
  \textbf{\bibinfo{volume}{8}}, \bibinfo{pages}{145--152},
  \doiprefix\url{https://doi.org/10.1038/nphoton.2013.343}
  (\bibinfo{year}{2013}).

\bibitem{Yi_2015}
\bibinfo{author}{Yi, X.}, \bibinfo{author}{Yang, Q.-F.}, \bibinfo{author}{Yang,
  K.~Y.}, \bibinfo{author}{Suh, M.-G.} \& \bibinfo{author}{Vahala, K.}
\newblock \bibinfo{journal}{\bibinfo{title}{Soliton frequency comb at microwave
  rates in a high-q silica microresonator}}.
\newblock {\emph{\JournalTitle{Optica}}} \textbf{\bibinfo{volume}{2}},
  \bibinfo{pages}{1078},
  \doiprefix\url{https://doi.org/10.1364/OPTICA.2.001078}
  (\bibinfo{year}{2015}).

\bibitem{Brasch_2016}
\bibinfo{author}{Brasch, V.} \emph{et~al.}
\newblock \bibinfo{journal}{\bibinfo{title}{Photonic chip{\textendash}based
  optical frequency comb using soliton cherenkov radiation}}.
\newblock {\emph{\JournalTitle{Science}}} \textbf{\bibinfo{volume}{351}},
  \bibinfo{pages}{357--360},
  \doiprefix\url{https://doi.org/10.1126/science.aad4811}
  (\bibinfo{year}{2016}).

\bibitem{Obrzud_2017}
\bibinfo{author}{Obrzud, E.}, \bibinfo{author}{Lecomte, S.} \&
  \bibinfo{author}{Herr, T.}
\newblock \bibinfo{journal}{\bibinfo{title}{Temporal solitons in
  microresonators driven by optical pulses}}.
\newblock {\emph{\JournalTitle{Nature Photonics}}}
  \textbf{\bibinfo{volume}{11}}, \bibinfo{pages}{600--607},
  \doiprefix\url{https://doi.org/10.1038/nphoton.2017.140}
  (\bibinfo{year}{2017}).

\bibitem{herrmann_theory_1994}
\bibinfo{author}{Herrmann, J.}
\newblock \bibinfo{journal}{\bibinfo{title}{Theory of {Kerr}-lens mode-locking:
  role of self-focusing and radially varying gain}}.
\newblock {\emph{\JournalTitle{J. Opt. Soc. Am. B}}}  (\bibinfo{year}{1994}).

\bibitem{t_brabec_hard-aperture_1993}
\bibinfo{author}{Brabec, T.}, \bibinfo{author}{Schmidt, A.~J.},
  \bibinfo{author}{Curley, P.~F.}, \bibinfo{author}{Spielmann, C.} \&
  \bibinfo{author}{Wintner, E.}
\newblock \bibinfo{journal}{\bibinfo{title}{Hard-aperture {Kerr}-lens mode
  locking}}.
\newblock {\emph{\JournalTitle{J. Opt. Soc. Am. B}}}
  \textbf{\bibinfo{volume}{10}} (\bibinfo{year}{1993}).

\bibitem{kurtner_mode-locking_1998}
\bibinfo{author}{Kurtner, F.~X.}, \bibinfo{author}{Au, J. A.~d.} \&
  \bibinfo{author}{Keller, U.}
\newblock \bibinfo{journal}{\bibinfo{title}{Mode-locking with slow and fast
  saturable absorbers-what's the difference?}}
\newblock {\emph{\JournalTitle{IEEE Journal of Selected Topics in Quantum
  Electronics}}} \textbf{\bibinfo{volume}{4}}, \bibinfo{pages}{159--168},
  \doiprefix\url{10.1109/2944.686719} (\bibinfo{year}{1998}).
\newblock \bibinfo{note}{Publisher: Institute of Electrical and Electronics
  Engineers (IEEE)}.

\bibitem{matsko_mode-locked_2011}
\bibinfo{author}{Matsko, A.~B.} \emph{et~al.}
\newblock \bibinfo{journal}{\bibinfo{title}{Mode-locked {Kerr} frequency
  combs}}.
\newblock {\emph{\JournalTitle{Optics Letters}}} \textbf{\bibinfo{volume}{36}},
  \bibinfo{pages}{2845}, \doiprefix\url{10.1364/ol.36.002845}
  (\bibinfo{year}{2011}).
\newblock \bibinfo{note}{Publisher: The Optical Society}.

\bibitem{haus_theory_1975}
\bibinfo{author}{Haus, H.~A.}
\newblock \bibinfo{journal}{\bibinfo{title}{Theory of mode locking with a fast
  saturable absorber}}.
\newblock {\emph{\JournalTitle{J. Appl. Phys.}}} \textbf{\bibinfo{volume}{46}}
  (\bibinfo{year}{1975}).

\bibitem{Liu2019}
\bibinfo{author}{Liu, X.}, \bibinfo{author}{Popa, D.} \&
  \bibinfo{author}{Akhmediev, N.}
\newblock \bibinfo{journal}{\bibinfo{title}{Revealing the transition dynamics
  from $q$ switching to mode locking in a soliton laser}}.
\newblock {\emph{\JournalTitle{Phys. Rev. Lett.}}}
  \textbf{\bibinfo{volume}{123}}, \bibinfo{pages}{093901},
  \doiprefix\url{10.1103/PhysRevLett.123.093901} (\bibinfo{year}{2019}).

\bibitem{Jung1995}
\bibinfo{author}{Jung, I.~D.}, \bibinfo{author}{K\"{a}rtner, F.~X.},
  \bibinfo{author}{Brovelli, L.~R.}, \bibinfo{author}{Kamp, M.} \&
  \bibinfo{author}{Keller, U.}
\newblock \bibinfo{journal}{\bibinfo{title}{Experimental verification of
  soliton mode locking using only a slow saturable absorber}}.
\newblock {\emph{\JournalTitle{Opt. Lett.}}} \textbf{\bibinfo{volume}{20}},
  \bibinfo{pages}{1892--1894}, \doiprefix\url{10.1364/OL.20.001892}
  (\bibinfo{year}{1995}).

\bibitem{ippen_principles_1994}
\bibinfo{author}{Ippen, E.~P.}
\newblock \bibinfo{journal}{\bibinfo{title}{Principles of passive mode
  locking}}.
\newblock {\emph{\JournalTitle{Applied Physics B Laser and Optics}}}
  \textbf{\bibinfo{volume}{58}}, \bibinfo{pages}{159--170},
  \doiprefix\url{10.1007/bf01081309} (\bibinfo{year}{1994}).
\newblock \bibinfo{note}{Publisher: Springer Science and Business Media LLC}.

\bibitem{Kartner1996}
\bibinfo{author}{Kartner, F.}, \bibinfo{author}{Jung, I.} \&
  \bibinfo{author}{Keller, U.}
\newblock \bibinfo{journal}{\bibinfo{title}{Soliton mode-locking with saturable
  absorbers}}.
\newblock {\emph{\JournalTitle{IEEE Journal of Selected Topics in Quantum
  Electronics}}} \textbf{\bibinfo{volume}{2}}, \bibinfo{pages}{540--556},
  \doiprefix\url{10.1109/2944.571754} (\bibinfo{year}{1996}).

\bibitem{Haus1996}
\bibinfo{author}{Haus, H.}, \bibinfo{author}{Jones, D.},
  \bibinfo{author}{Ippen, E.} \& \bibinfo{author}{Wong, W.}
\newblock \bibinfo{journal}{\bibinfo{title}{Theory of soliton stability in
  asynchronous modelocking}}.
\newblock {\emph{\JournalTitle{Journal of Lightwave Technology}}}
  \textbf{\bibinfo{volume}{14}}, \bibinfo{pages}{622--627},
  \doiprefix\url{10.1109/50.491401} (\bibinfo{year}{1996}).

\bibitem{siegman_lasers_1986}
\bibinfo{author}{Siegman, A.~E.}
\newblock \bibinfo{title}{Lasers} (\bibinfo{publisher}{University Science
  Books}, \bibinfo{year}{1986}).
\newblock \bibinfo{note}{Section: 20}.

\bibitem{coen_modeling_2012}
\bibinfo{author}{Coen, S.}, \bibinfo{author}{Randle, H.~G.},
  \bibinfo{author}{Sylvestre, T.} \& \bibinfo{author}{Erkintalo, M.}
\newblock \bibinfo{journal}{\bibinfo{title}{Modeling of octave-spanning {Kerr}
  frequency combs using a generalized mean-field {Lugiato}–{Lefever} model}}.
\newblock {\emph{\JournalTitle{Optics Letters}}} \textbf{\bibinfo{volume}{38}},
  \bibinfo{pages}{37}, \doiprefix\url{10.1364/ol.38.000037}
  (\bibinfo{year}{2012}).
\newblock \bibinfo{note}{Publisher: The Optical Society}.

\bibitem{meller_mode-locking_2017}
\bibinfo{author}{Meller, M.~E.}, \bibinfo{author}{Yefet, S.} \&
  \bibinfo{author}{Pe'er, A.}
\newblock \bibinfo{journal}{\bibinfo{title}{Mode-{Locking} {With} {Ultra}-{Low}
  {Intra}-{Cavity} {Pulse} {Intensity} {Using} {Enhanced} {Kerr}
  {Nonlinearity}}}.
\newblock {\emph{\JournalTitle{IEEE Journal of Quantum Electronics}}}
  \textbf{\bibinfo{volume}{53}}, \doiprefix\url{10.1109/jqe.2017.2670544}
  (\bibinfo{year}{2017}).

\bibitem{dunlop_master_1997}
\bibinfo{author}{Dunlop, A.}, \bibinfo{author}{Firth, W.} \&
  \bibinfo{author}{Wright, E.}
\newblock \bibinfo{journal}{\bibinfo{title}{Master equation for spatio-temporal
  beam propagation and {Kerr} lens mode-locking}}.
\newblock {\emph{\JournalTitle{Optics Communications}}}
  \textbf{\bibinfo{volume}{138}}, \bibinfo{pages}{211--226},
  \doiprefix\url{10.1016/S0030-4018(97)00037-0} (\bibinfo{year}{1997}).

\bibitem{c_j_chen_self-starting_1995}
\bibinfo{author}{C.~J.~Chen, P. K. A.~W.} \& \bibinfo{author}{Menyuk, C.~R.}
\newblock \bibinfo{journal}{\bibinfo{title}{Self-starting of passively
  mode-locked lasers with fast saturable absorbers}}.
\newblock {\emph{\JournalTitle{Optics Letters}}} \textbf{\bibinfo{volume}{20}}
  (\bibinfo{year}{1995}).

\bibitem{parshani_diffractive_2021}
\bibinfo{author}{Parshani, I.}, \bibinfo{author}{Bello, L.},
  \bibinfo{author}{Meller, M.-E.} \& \bibinfo{author}{Pe’er, A.}
\newblock \bibinfo{journal}{\bibinfo{title}{Diffractive saturable loss
  mechanism in {Kerr}-lens mode-locked lasers: direct observation and
  simulation}}.
\newblock {\emph{\JournalTitle{Optics Letters}}} \textbf{\bibinfo{volume}{46}},
  \bibinfo{pages}{1530}, \doiprefix\url{10.1364/OL.418788}
  (\bibinfo{year}{2021}).

\bibitem{xu_spontaneous_2021}
\bibinfo{author}{Xu, G.} \emph{et~al.}
\newblock \bibinfo{journal}{\bibinfo{title}{Spontaneous symmetry breaking of
  dissipative optical solitons in a two-component {Kerr} resonator}}.
\newblock {\emph{\JournalTitle{Nature Communications}}}
  \textbf{\bibinfo{volume}{12}}, \bibinfo{pages}{4023},
  \doiprefix\url{10.1038/s41467-021-24251-0} (\bibinfo{year}{2021}).

\bibitem{del_bino_symmetry_2017}
\bibinfo{author}{Del~Bino, L.}, \bibinfo{author}{Silver, J.~M.},
  \bibinfo{author}{Stebbings, S.~L.} \& \bibinfo{author}{Del'Haye, P.}
\newblock \bibinfo{journal}{\bibinfo{title}{Symmetry {Breaking} of
  {Counter}-{Propagating} {Light} in a {Nonlinear} {Resonator}}}.
\newblock {\emph{\JournalTitle{Scientific Reports}}}
  \textbf{\bibinfo{volume}{7}}, \bibinfo{pages}{43142},
  \doiprefix\url{10.1038/srep43142} (\bibinfo{year}{2017}).

\bibitem{hendry_spontaneous_2018}
\bibinfo{author}{Hendry, I.} \emph{et~al.}
\newblock \bibinfo{journal}{\bibinfo{title}{Spontaneous symmetry breaking and
  trapping of temporal {Kerr} cavity solitons by pulsed or amplitude-modulated
  driving fields}}.
\newblock {\emph{\JournalTitle{Physical Review A}}}
  \textbf{\bibinfo{volume}{97}}, \bibinfo{pages}{053834},
  \doiprefix\url{10.1103/PhysRevA.97.053834} (\bibinfo{year}{2018}).

\bibitem{SIGLER2005305}
\bibinfo{author}{Sigler, A.} \& \bibinfo{author}{Malomed, B.~A.}
\newblock \bibinfo{journal}{\bibinfo{title}{Solitary pulses in linearly coupled
  cubic–quintic ginzburg–landau equations}}.
\newblock {\emph{\JournalTitle{Physica D: Nonlinear Phenomena}}}
  \textbf{\bibinfo{volume}{212}}, \bibinfo{pages}{305--316},
  \doiprefix\url{https://doi.org/10.1016/j.physd.2005.10.004}
  (\bibinfo{year}{2005}).

\bibitem{akhmediev1993novel}
\bibinfo{author}{Akhmediev, N.} \& \bibinfo{author}{Ankiewicz, A.}
\newblock \bibinfo{journal}{\bibinfo{title}{Novel soliton states and
  bifurcation phenomena in nonlinear fiber couplers}}.
\newblock {\emph{\JournalTitle{Physical review letters}}}
  \textbf{\bibinfo{volume}{70}}, \bibinfo{pages}{2395} (\bibinfo{year}{1993}).

\bibitem{chu1993soliton}
\bibinfo{author}{Chu, P.~L.}, \bibinfo{author}{Malomed, B.~A.} \&
  \bibinfo{author}{Peng, G.-D.}
\newblock \bibinfo{journal}{\bibinfo{title}{Soliton switching and propagation
  in nonlinear fiber couplers: analytical results}}.
\newblock {\emph{\JournalTitle{JOSA B}}} \textbf{\bibinfo{volume}{10}},
  \bibinfo{pages}{1379--1385} (\bibinfo{year}{1993}).

\bibitem{nguyen2020reversible}
\bibinfo{author}{Nguyen, V.~H.} \emph{et~al.}
\newblock \bibinfo{journal}{\bibinfo{title}{Reversible ultrafast soliton
  switching in dual-core highly nonlinear optical fibers}}.
\newblock {\emph{\JournalTitle{Optics Letters}}} \textbf{\bibinfo{volume}{45}},
  \bibinfo{pages}{5221--5224} (\bibinfo{year}{2020}).

\bibitem{MAGNI1993348}
\bibinfo{author}{Magni, V.}, \bibinfo{author}{Cerullo, G.} \&
  \bibinfo{author}{{De Silvestri}, S.}
\newblock \bibinfo{journal}{\bibinfo{title}{Abcd matrix analysis of propagation
  of gaussian beams through kerr media}}.
\newblock {\emph{\JournalTitle{Optics Communications}}}
  \textbf{\bibinfo{volume}{96}}, \bibinfo{pages}{348--355},
  \doiprefix\url{https://doi.org/10.1016/0030-4018(93)90284-C}
  (\bibinfo{year}{1993}).

\bibitem{salin_mode_1991}
\bibinfo{author}{Salin, F.}, \bibinfo{author}{Piché, M.} \&
  \bibinfo{author}{Squier, J.}
\newblock \bibinfo{journal}{\bibinfo{title}{Mode locking of {Ti}:{Al}\_2o\_3
  lasers and self-focusing: a {Gaussian} approximation}}.
\newblock {\emph{\JournalTitle{Optics Letters}}} \textbf{\bibinfo{volume}{16}},
  \bibinfo{pages}{1674}, \doiprefix\url{10.1364/ol.16.001674}
  (\bibinfo{year}{1991}).
\newblock \bibinfo{note}{Publisher: The Optical Society}.

\bibitem{yoo_byung_duk_numerical_2005}
\bibinfo{author}{Yoo, B.~D.} \emph{et~al.}
\newblock \bibinfo{journal}{\bibinfo{title}{Numerical {Analysis} of
  {Soft}-{Aperture} {Kerr}-{Lens} {Mode} {Locking} in {Ti}:{Sapphire} {Laser}
  {Cavities} by {Using} {Nonlinear} {ABCD} {Matrices}}}.
\newblock {\emph{\JournalTitle{Journal of the Korean Physical Society}}}
  \textbf{\bibinfo{volume}{46}}, \bibinfo{pages}{1131--1136}
  (\bibinfo{year}{2005}).

\bibitem{Juang1997}
\bibinfo{author}{Juang, D.-G.}, \bibinfo{author}{Chen, Y.-C.},
  \bibinfo{author}{Hsu, S.-H.}, \bibinfo{author}{Lin, K.-H.} \&
  \bibinfo{author}{Hsieh, W.-F.}
\newblock \bibinfo{journal}{\bibinfo{title}{Differential gain and buildup
  dynamics of self-starting kerr lens mode-locked ti:sapphire laser without an
  internal aperture}}.
\newblock {\emph{\JournalTitle{J. Opt. Soc. Am. B}}}
  \textbf{\bibinfo{volume}{14}}, \bibinfo{pages}{2116--2121},
  \doiprefix\url{10.1364/JOSAB.14.002116} (\bibinfo{year}{1997}).

\bibitem{Cerullo1994}
\bibinfo{author}{Cerullo, G.}, \bibinfo{author}{Silvestri, S.~D.} \&
  \bibinfo{author}{Magni, V.}
\newblock \bibinfo{journal}{\bibinfo{title}{Self-starting kerr-lens mode
  locking of a ti:sapphire laser}}.
\newblock {\emph{\JournalTitle{Opt. Lett.}}} \textbf{\bibinfo{volume}{19}},
  \bibinfo{pages}{1040--1042}, \doiprefix\url{10.1364/OL.19.001040}
  (\bibinfo{year}{1994}).

\bibitem{Henrich1997}
\bibinfo{author}{Henrich, B.} \& \bibinfo{author}{Beigang, R.}
\newblock \bibinfo{journal}{\bibinfo{title}{Self-starting kerr-lens mode
  locking of a nd:yag-laser}}.
\newblock {\emph{\JournalTitle{Optics Communications}}}
  \textbf{\bibinfo{volume}{135}}, \bibinfo{pages}{300--304},
  \doiprefix\url{https://doi.org/10.1016/S0030-4018(96)00673-6}
  (\bibinfo{year}{1997}).

\bibitem{yefet_mode_2013}
\bibinfo{author}{Yefet, S.} \& \bibinfo{author}{Pe'er, A.}
\newblock \bibinfo{journal}{\bibinfo{title}{Mode locking with enhanced
  nonlinearity - a detailed study}}.
\newblock {\emph{\JournalTitle{Optics Express}}} \textbf{\bibinfo{volume}{21}},
  \bibinfo{pages}{19040}, \doiprefix\url{10.1364/oe.21.019040}
  (\bibinfo{year}{2013}).
\newblock \bibinfo{note}{Publisher: The Optical Society}.

\bibitem{yefet_review_2013}
\bibinfo{author}{Yefet, S.} \& \bibinfo{author}{Pe'er, A.}
\newblock \bibinfo{journal}{\bibinfo{title}{A {Review} of {Cavity} {Design} for
  {Kerr} {Lens} {Mode}-{Locked} {Solid}-{State} {Lasers}}}.
\newblock {\emph{\JournalTitle{Applied Sciences}}}
  \textbf{\bibinfo{volume}{3}}, \bibinfo{pages}{694--724},
  \doiprefix\url{10.3390/app3040694} (\bibinfo{year}{2013}).

\bibitem{rp_photonics_sat_power}
\bibinfo{howpublished}{https://www.rp-photonics.com/saturation\_power.html}.

\end{thebibliography}

\end{document}